 \documentclass[twocolumn, prl, showpacs]{revtex4}
 \newif\ifGALLEYversion\GALLEYversionfalse

\usepackage{ifthen}
\usepackage{graphicx}
\usepackage{amsmath}

\usepackage{color}

\ifthenelse{\boolean{GALLEYversion}}{
    \marginparwidth 2.7in
    \marginparsep 0.5in
    \def\abm#1{\marginpar{\small AB: #1}}
    \def\thm#1{\marginpar{\small HT: #1}}
    \def\rdm#1{\marginpar{\small RD: #1}}
    \def\akm#1{\marginpar{\small AK: #1}}
}{
    \def\abm#1{\relax}
    \def\thm#1{\relax}
    \def\rdm#1{\relax}
    \def\akm#1{\relax}
}

\usepackage{graphicx}

\begin{document}

\title{New superconducting and semiconducting Fe-B compounds predicted with an {\it ab initio} evolutionary search}

\author{A.N. Kolmogorov,$^{1}$ S. Shah,$^{1}$ E.R. Margine,$^{1}$ A.F. Bialon,$^{2}$ T. Hammerschmidt,$^{2}$ and R. Drautz$^{2}$}

\affiliation{$^{1}$Department of Materials, University of Oxford, Parks Road, Oxford OX1 3PH, United Kingdom}
\affiliation{$^{2}$Atomistic Modelling and Simulation, ICAMS, Ruhr-Universit\"at Bochum, D-44801 Bochum, Germany}

\date{\today}

\begin{abstract}
  {New candidate ground states at 1:4, 1:2, and 1:1 compositions are
  identified in the well known Fe-B system via a combination of {\it
  ab initio} high-throughput and evolutionary searches. We show that
  the proposed oP12-FeB$_2$ stabilizes by a break up of 2D boron
  layers into 1D chains while oP10-FeB$_4$ stabilizes by a distortion
  of a 3D boron network. The uniqueness of these configurations gives
  rise to a set of remarkable properties: oP12-FeB$_2$ is expected to
  be the first semiconducting metal diboride and oP10-FeB$_4$ is shown
  to have the potential for phonon-mediated superconductivity with a
  $T_c$ of 15-20 K.}
\end{abstract}



\maketitle



A range of advanced compound prediction methods has been developed
recently to accelerate the experimental search for materials
displaying novel physics or technologically relevant
features \cite{Woodley,predict,Oganov,Hautier}. Unconstrained
structural optimization with evolutionary algorithms (EAs) has shown
the ability to predict complex configurations given only the
composition leading to identification of exotic high-pressure
phases \cite{Oganov}. High-throughput screening with data mining
techniques has proven effective in revealing compositions favorable
to form in large sets of multi-component systems \cite{Hautier}. In
this study we demonstrate that new ambient-pressure materials with
appealing properties could be found in such a well-known and
accessible binary system as Fe-B.

The experimental research on Fe-B compounds has been driven primarily
by their potential to serve as a hardening agent in steels
\cite{prec1} or as hard protective coatings
\cite{Coatings1,Coatings2}. According to the latest experimental phase
diagram \cite{FeB-PhD}, FeB and Fe$_2$B are the only reproducible low
temperature phases that have been shown to crystallize in the oP8 (or
the related oS8 \cite{CrB_pure_and_CrB_inter}) and tI12
configurations, respectively. Less is known about the boron-rich
ordered phases with only a few reports available: observation of a
metastable FeB$_{49}$ intercalation compound \cite{FeB49} and possible
synthesis of amorphous \cite{aFeB2} and the AlB$_2$-type \cite{FeB2}
iron diborides. Previous modelling work on Fe-B compounds has given
insights into their binding, magnetic, and structural properties
\cite{TMB2,TMB4,petTMB,TM2B,manyFeB,Fe23B6} but has not systematically
explored the possibility of obtaining new stable iron borides.

Our reexamination of the Fe-B system within density functional theory
(DFT) begins with a high-throughput scan of known configurations
listed in the Inorganic Crystal Structure Database ICSD
\cite{ICSD}. We show that never observed oP6-FeB$_2$ (hP6-FeB$_2$) and
tI16-FeB phases are marginally stable relative to the known
compounds. The proposed Fe-B ground states are then refined with an
{\it ab initio} evolutionary search that suggests oP10-FeB$_4$ and
oP12-FeB$_2$ to be ground states at 1:4 and 1:2 compositions. The
prediction of the brand new stable structure types is surprising as
transition metal (TM) borides tend to crystallize in configurations
correlating well between the $3d$, $4d$, and $5d$ series
\cite{MB_review}. We link the stabilization of the Fe-B phases to the
structural changes in the B networks that lead to radically new
properties. At 1:2 metal-boron composition, famous for the outstanding
MgB$_2$ superconductor \cite{MgB2_exp} and the hardest metal-based
ReB$_2$ material \cite{ReB2}, oP12-FeB$_2$ stands out as the first
semiconducting metal diboride made out of B chains rather than B
layers. At 1:4 composition, the nonmagnetic oP10-FeB$_4$ is examined
using electron-phonon ($e$-ph) calculations and predicted to be,
subject to spin fluctuation effects \cite{spin}, a superconductor with
an unexpectedly high $T_c$ of 15-20~K. The critical temperature falls
between the typical 10~K $T_c$ of TM borides \cite{MoB4} and the 39~K
$T_c$ of MgB$_2$ \cite{MgB2_exp}. If synthesized, oP10-FeB$_4$ could
extend the family of recently discovered iron-based
LaFeAsO$_{1-x}$F$_x$ and FeSe superconducting materials
\cite{Fe-based} but have the conventional phonon-mediated coupling
mechanism.

We carry out the high-throughput scan by calculating formation
enthalpies at $T=0$~K and $P=0$~GPa with {\small VASP} \cite{VASP} for
over 40 commonly observed $M$-B and $M$-C ICSD structure
types \cite{SuppMat} in the whole composition range. The B-rich end
of the phase diagram is further explored with the Module for {\it Ab
Initio} Structure Evolution ({\small MAISE}) \cite{MAISE} linked with
{\small VASP} which enables an EA search for the lowest enthalpy
ordered phases. The unconstrained structural optimization is carried
out for most likely to occur 1:6, 1:4, 1:3, and 1:2 compositions
starting from random unit cells of up to 15 atoms (for further details
see supplementary material \cite{SuppMat}).

Finding ground states also depends on the accuracy of the simulation
method and the inclusion of important Gibbs energy contributions
\cite{Woodley}. We use the projector augmented waves method \cite{PAW}
and allow spin polarization unless stated otherwise; the chosen energy
cutoff of 500~eV and dense Monkhorst-Pack $k$-meshes
\cite{MONKHORST_PACK} ensure numerical convergence of formation energy
differences to typically 1-2 meV/atom. We employ the
Perdew-Burke-Ernzerhof (PBE) exchange-correlation (xc) functional
\cite{PBE} within the generalized gradient approximation (GGA) that
provides a realistic description of the Fe ground state
\cite{Fe}. Tests in the supplementary material \cite{SuppMat}
demonstrate independence of our key finding, the stability of new
phases at the 1:2 and 1:4 Fe-B compositions with respect to known
compounds, on the choice of the xc functional. The ground state of B
is modelled as $\alpha$-B which has been recently shown to be only 3-4
meV/atom above the more complex $\beta$-B in the 0-300~K temperature
range \cite{aB}.  We include phonon corrections to $G(T)$ using a
finite displacement method as implemented in PHON \cite{PHON}. The
$e$-ph calculations are carried out within the linear response theory
using the {\small Quantum-ESPRESSO} package
\cite{PWscf,kq_meshes}. Our tests show no magnetic moment in relevant
B-rich phases allowing us to do the EA, phonon, and $e$-ph simulations
without spin polarization.

\begin{figure}[t]
\begin{center}
\hspace{-0.5 cm}
\includegraphics[width=80mm,angle=0]{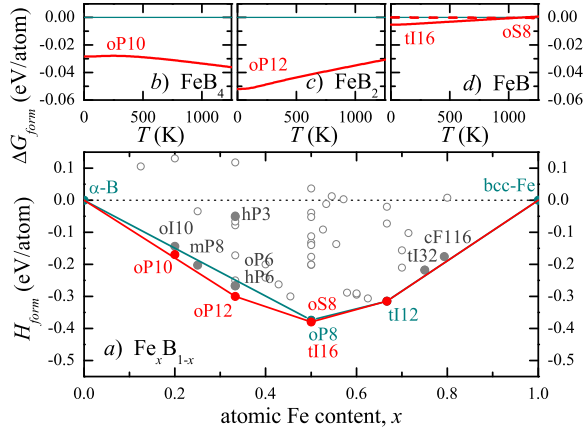}
\caption{ \small (Color online) Stability of Fe-B alloys calculated
with GGA-PBE: {\it a}) formation enthalpy; {\it b-d}) Gibbs energy with
thermodynamic corrections due to the vibrational entropy for selected
candidate phases w.r.t. the $\alpha$-B$\leftrightarrow$oP8-FeB
tie-line.}
\label{fig1}
\end{center}
\end{figure}
Figure 1(a) summarizes calculated $T=0$~K formation energies of the
considered Fe$_x$B$_{1-x}$ ordered structures with the convex hull
drawn (in cyan) through the known oP8-FeB and tI12-Fe$_2$B ground
states; other relevant oI10, oP10, mP8, hP3, hP6, oP6, oP12, oS8,
tI16, tI32 and cF116 structures correspond to the CrB$_4$, FeB$_4$
(proposed), FeB$_3$ (proposed), AlB$_2$, ReB$_2$, RuB$_2$, FeB$_2$
(proposed), CrB, MoB, Ni$_3$P, and Cr$_{23}$C$_{6}$ prototypes,
respectively. Phases with $x>0.5$ show an expected ordering, with
tI12-Fe$_2$B being stable and Fe$_3$B and Fe$_{23}$B$_6$ being
metastable by less than 20 meV/atom. For $x\leq0.5$, we find a set of
phases that are below or close to the
$\alpha$-B$\leftrightarrow$oP8-FeB tie-line to be viable ground state
candidates. We discuss their relative stability using structural and
electronic density of states (DOS) information shown in Figs. 2 and 3.

The similarity of the local coordinations in oP8, oS8, and tI16 at 1:1
composition was discussed previously in Ref. \cite{TMB}. In tI16 the B
chains extend in two directions, a feature that could differentiate
the structure's mechanical response to external load from the
behaviour of the other two polymorphs. Fig. 1(d) reflects the
difference in the vibrational properties of oP8 (oS8) and tI16 and
could explain why tI16-FeB, marginally the most stable phase at $T=0$
K in our calculations, has never been observed. The electronic DOS of
FeB in the three configurations are rather similar (Fig. 3): they are
all metallic and magnetic and have bonding $p$-$d$ hybridized states
in the $-6\,$-$\,-3$ eV energy range.

\begin{figure}[t]
\begin{center}
\includegraphics[height=60mm,angle=0]{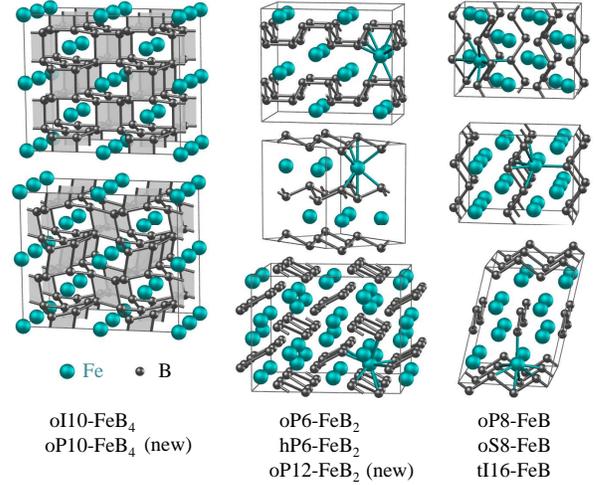}
\caption{ \small (Color online) Competing B-rich Fe-B phases; cell
parameters are given in the supplementary material \cite{SuppMat}.}
\label{fig2}
\end{center}
\end{figure}
At 1:2 composition, all known metal borides stable under normal
conditions are composed of 2D boron layers that are flat in hP3,
armchair in oP6, zigzag in hP6, or mixed in hR18 \cite{ICSD}. A
detailed rigid band approximation study of the TMB$_2$ phases linked
the distortion of the B layers to population of antibonding TM-TM and
TM-B orbitals in hP3-TMB$_2$ with high $d$-electron count \cite{TMB2}.
The projected DOS in hP3-FeB$_2$ (Fig. 3) shows a mismatch in the
maxima of the filled B and Fe states in the $-7\,$-$\,-3$ eV range and
a high DOS at the Fermi level resulting in a magnetic moment of 0.26
$\mu_B$/atom. The magnetisation energy of 11 meV/atom is insufficient
to stabilize the hP3-FeB phase which leaves it 200 meV/atom above the
$\alpha$-B$\leftrightarrow$oP8-FeB tie-line. Puckering of the B layers
in non-magnetic oP6 and hP6 proves to be a more favorable way of
reducing the high DOS at the Fermi level: Fig. 3 shows a higher
hybridization of the B-$p$ and Fe-$d$ states with the antibonding
$p-d$ states now lying just above $E_F$. The net result of the more
bimodal shape of the DOS is a 217 meV/atom gain in stability. Even
more dramatic structural and electronic changes take place in
oP12-FeB$_2$ discovered in our EA search. The disintegration of the B
layers opens up a $\sim0.5$ eV band gap (likely underestimated in our
semilocal DFT treatment \cite{bandgap}) and leads to an additional 33
meV/atom gain in stability. This finding rules out the existence of
hP3-FeB$_2$ that has been a subject of
controversy \cite{FeB2_controversy}; oP12-FeB$_2$ is below the
$\alpha$-B$\leftrightarrow$oP8-FeB tie-line by over 30 meV/atom
(Fig. 1(c)) in the whole $T$ range and should be synthesizable.

At 1:3 composition, the EA suggests a new mP8 phase \cite{SuppMat} that
breaks the $\alpha$-B$\leftrightarrow$oP8-FeB tie-line by 15 meV/atom;
however it is found to be metastable w.r.t. $\alpha$-B and
oP12-FeB$_2$ at all temperatures.

\begin{figure}[t]
\begin{center}
\includegraphics[width=68mm,angle=0]{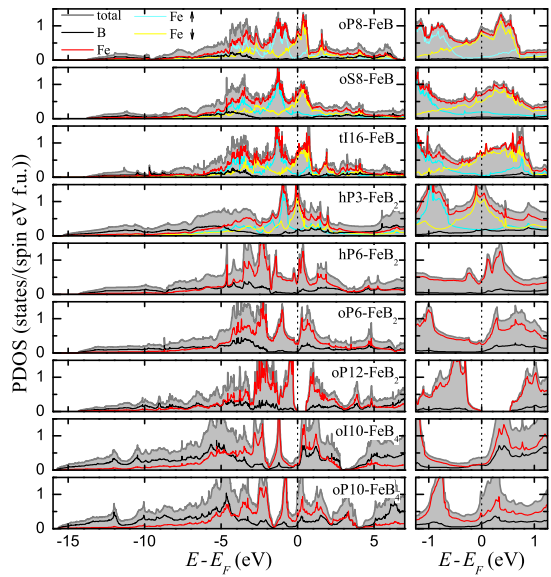}
\caption{ \small (Color online) Calculated density of states in selected
iron borides (the lower five compounds are non-magnetic).}
\label{fig3}
\end{center}
\end{figure}
%
At 1:4 composition, the observed proximity of the oI10-FeB$_4$ phase
to the $\alpha$-B$\leftrightarrow$oP8-FeB tie-line is intriguing as
there have been explicit references to unsuccessful attempts to
synthesize this phase \cite{TMB4}. Unexpectedly, a phonon dispersion
calculation showed dynamical instability of oI10-FeB$_4$, with
imaginary frequencies reaching $208i$ cm$^{-1}$ for a $\Gamma$-point
phonon in the conventional 10-atom unit cell. Using the phonon
eigenvector that skews the rectangle B building units in the $x$-$y$
plane into parallelograms we have constructed a new structure type,
oP10-FeB$_4$. The considerable energy gain of 28 meV/atom and no
imaginary frequencies in the phonon spectrum make oP10-FeB$_4$
thermodynamically and dynamically stable in the considered temperature
range relative to known phases [Fig. 1(b)]; with phonon corrections
included oP10-FeB$_4$ lies 3 meV/atom above the
$\alpha$-B$\leftrightarrow$oP12-FeB$_2$ tie-line at $T=0$~K but 10
meV/atom below the tie-line at $T=900$~K. Additional no-symmetry
relaxations of distorted oI10 and oP10 supercells with 10, 20, and 40
atoms have consistently produced oP10-FeB$_4$ as the most stable
configuration. The EA search has also shown that Fe$_2$B$_8$ cells
converge to oP10-FeB$_4$ while Fe$_3$B$_{12}$ cells evolve into a new
mS30-FeB$_4$ phase metastable by 6 meV/atom. The tests seemed
necessary due to a counterintuitive evolution of the DOS in the oI10
to oP10 transformation: the Fermi level in oP10-FeB$_4$ catches the
edge of the antibonding $p_{x,y}$-$d_{x^2-y^2}$ peak resulting in a
high $n(E_F)=1.0$ states/(eV spin f.u.); the feature is unusual as
stable compounds tend to have the Fermi level lying in the pseudogap
\cite{LiB2}.

\begin{figure}[t]
\begin{center}
\includegraphics[width=70mm,angle=0]{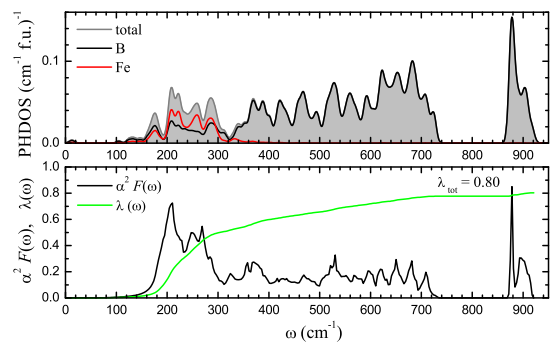}
\caption{ \small (Color online) Top: total and projected phonon
density of states (PHDOS) in oP10-FeB$_4$. Bottom: Eliashberg function
and the strength of the electron-phonon coupling.}
\label{fig4}
\end{center}
\end{figure}
The naturally electron-doped oP10-FeB$_4$ candidate material with
strong covalent bonds is next analyzed for superconducting
features. We use the linear response theory and fine $k$- and
$q$-meshes \cite{kq_meshes,SuppMat} to calculate the phonon DOS
(PHDOS), Eliashberg function ($\alpha^2F(\omega)$), and strength of
the $e$-ph coupling ($\lambda(\omega)$). The phonon spectrum in Fig. 4
can be divided into three regions with mixed Fe-B modes (0-320
cm$^{-1}$), B modes with a relatively flat PHDOS (320-740 cm$^{-1}$),
and B modes involving in-plane optical distortions of B parallelograms
(860-920 cm$^{-1}$). The Eliashberg function integrates to a large
$\lambda_{{\text {tot}}}=0.80$ and gives the logarithmic average
$\left<\omega\right>_{ln}=430$ cm$^{-1}$. While key contributions to
$\lambda_{{\text {tot}}}\sim0.8$ in CaC$_6$ and MgB$_2$ come from the
low-frequency Ca modes ($\omega<150$ cm$^{-1}$) \cite{CaC6} and the
high-frequency B modes (500 cm$^{-1}<\omega<560$ cm$^{-1}$)
\cite{MgB2_Liu}, respectively, nearly 60\% of $\lambda_{{\text
{tot}}}$ in oP10-FeB$_4$ is generated by the mixed Fe-B modes in the
160-300 cm$^{-1}$ range. $\left<\omega\right>_{ln}$ in oP10-FeB$_4$ is
found to be much closer to the MgB$_2$ value of $\sim450$ cm$^{-1}$
\cite{MgB2_Liu}, rather than the CaC$_6$ value of $\sim200$ cm$^{-1}$
\cite{CaC6}. Using the Allen-Dynes formula \cite{Allen-Dynes} and
typical $\mu^*$ of 0.14-0.10 we estimate the $T_c$ in oP10-FeB$_4$ to
be 15-20~K. The compound has two 3D Fermi surfaces centered at
$\Gamma$ and R points and the $T_c$ may be further enhanced by the
multiband effect. Because of the large gradient of the DOS near the
Fermi level the superconducting properties may be strongly affected by
the presence of vacancies or impurities. Although oP10-FeB$_4$ is
found to have neither ferro- nor antiferromagnetic moment, spin
fluctuations \cite{spin} could play a critical role in the pairing
mechanism and should be examined carefully using input from
experiment.

In summary, our search for new compounds in the common Fe-B system
demonstrates the necessity to go beyond standard structure types: The
EA-driven unconstrained structural optimization has uncovered a set of
viable ground states, with new oP10-FeB$_4$ and oP12-FeB$_2$ shown to
be thermodynamically stable by over 25 meV/atom relative to the known
$\alpha$-B and oP8-FeB. To the best of our knowledge, the identified
boron-rich phases have been never observed before and their discovery
may require finding suitable kinetic routes. The presented analysis of
the structural and electronic properties shows how the phases
stabilize and what new physics they are expected to exhibit if
synthesized. (i) oP10-FeB$_4$ could become yet another exception to
Matthias' rules \cite{Pickett} that recommend staying away from
magnetic elements when designing new superconductors. This compound
has a high DOS at the Fermi level leading to $\lambda=0.80$
and a surprisingly high $T_c$ of 15-20~K. (ii) oP12-FeB$_2$ is
predicted to be the first metal diboride semiconductor with a
$\sim0.5$~eV band gap (time-dependent DFT or Green's functions
techniques will likely give a larger value \cite{bandgap}). (iii) The
proposed materials may also exhibit appealing mechanical properties as
hardness tends to be higher for $M_x$B$_{1-x}$ with
$x<0.5$ \cite{ReB2}.

ANK and SS acknowledge the support of the EPSRC and the Oxford
Supercomputing Centre. AFB, TH and RD acknowledge financial support
through ThyssenKrupp AG, Bayer MaterialScience AG, Salzgitter
Mannesmann Forschung GmbH, Robert Bosch GmbH, Benteler Stahl/Rohr
GmbH, Bayer Technology Services GmbH and the state of North-Rhine
Westphalia as well as the EU in the framework of the ERDF.



\begin{thebibliography}{99}
\bibliographystyle{ieeetr}

\bibitem{Woodley}
S.W. Woodley and R. Catlow, Nature Mater., {\bf 7}, 937 (2008).

\bibitem{predict}
R. Marto\v{n}\'{a}k, A. Laio, and M. Parrinello, Phys. Rev. Lett. {\bf
90}, 075503 (2003); S. Curtarolo {\it et al.}, Phys. Rev. Lett. {\bf
91}, 135503 (2003); C.C. Fischer {\it et al.}, Nat. Mater. {\bf 5},
641 (2006); G. Trimarchi, A.J. Freeman, and A. Zunger, Phys. Rev. B
{\bf 80}, 092101 (2009); C.J. Pickard and R.J. Needs,
Phys. Rev. Lett. {\bf 97}, 045504 (2006).

\bibitem{Oganov}
A.R. Oganov and C.W. Glass, J. Phys.: Condens. Matter {\bf 20}, 064210 (2008);
A.R. Oganov {\it et al.}, Nature (London) {\bf 457}, 863 (2009).

\bibitem{Hautier}
G. Hautier {\it et al.}, Chem. Mater. {\bf 22}, 3762 (2010).


\bibitem{prec1}
L. Lanier, G. Metauer, and M. Moukassi, Mikrochim. Acta {\bf 114/115},
353 (1994);
S. Watanabe, H. Ohtani, and T. Kunitake, Transactions ISIJ {\bf 23}, 120 (1983).



\bibitem{Coatings1}
I.~Campos {\it et al.}, Mater. Sci. Eng. {\bf A352}, 261 (2003).

\bibitem{Coatings2}
S. Sen, U. Sen, and C. Bindal, Vacuum {\bf 77}, 195 (2005).

\bibitem{FeB-PhD}
T.~van~Rompaey, K.C.~Hari~Kumar, and P.~Wollants, J. Alloys Compd. {\bf 334}, 17
3 (2002).

\bibitem{CrB_pure_and_CrB_inter}
V.A.~Barinov {\it et al.}, Phys. Stat. Sol. A {\bf 123}, 527 (1991);
T.~Kanaizuka, Phys. Stat. Sol. A {\bf 69}, 739 (1982).

\bibitem{FeB49}
K. Balani, A. Agarwala, and N.B. Dahotre, J. Appl. Phys. {\bf 99}, 044904 (2006).

\bibitem{aFeB2}
K.~Moorjani {\it et al.}, J. Appl. Phys {\bf 57}, 3445 (1985).

\bibitem{FeB2}
L.G.~Voroshnin {\it et al.}, Met. Sci. Heat Treat. [Metal. i Term. Obrabotka Metal.] {\bf 12}, 732 (1970).

\bibitem{TMB2} 
J.K. Burdett, E. Canadell, and G.J. Miller, J. Am. Chem. Soc. {\bf 108}, 6561 (1986).

\bibitem{TMB4}
J.K. Burdett and E. Canadell, Inorg. Chem. {\bf 27}, 4437 (1988).

\bibitem{petTMB}
P.~Mohn and D.G.~Pettifor, J. Phys. C: Solid State Phys. {\bf 21}, 2829 (1988).

\bibitem{TM2B}
P.~Mohn, J. Phys. C: Solid State Phys. {\bf 21}, 2841 (1988).

\bibitem{manyFeB}
W.Y.~Ching {\it et al.}, Phys. Rev. B {\bf 42}, 4460 (1990).

\bibitem{Fe23B6}
P.R.~Ohodnicki, Jr. {\it et al.}, Phys. Rev. B {\bf 78}, 144414 (2008).

\bibitem{ICSD}
G.~Bergerhoff and I.D.~Brown, in {\it Crystallographic Databases}, edited by F.H.~Allen {\it et al.} (International Union of Crystallography, Chester,1987).

\bibitem{MB_review}
J.J. Zuckerman and A.P. Hagen, (EDT), Inorganic Reactions And Methods (2007).

\bibitem{MgB2_exp}
J. Nagamatsu {\it el al.}, Nature (London) {\bf 410}, 63 (2001).

\bibitem{ReB2}
H.-Y. Chung {\it et al.}, Science {\bf 316}, 436 (2007).

\bibitem{spin}
I.I. Mazin {\it et al.}, Phys. Rev. Lett. {\bf 101}, 057003 (2008);
M. Wierzbowska, Eur. Phys. J. B {\bf 48}, 207 (2005). S.K. Bose
J. Phys.: Condens. Matter {\bf 21}, 025602 (2009).

\bibitem{MoB4}
J.W. Simonson {\it et al.}, J. Supercond. Nov. Magn. {\bf 23}, 417 (2010) and references therein.

\bibitem{Fe-based}
Y. Kamihara {\it et al.}, J. Am. Chem. Soc. {\bf 130}, 3296 (2008);
F.C. Hsu {\it et al.}, Proc. Natl. Acad. Sci. U.S.A. {\bf 105}, 14262
(2008); I.I. Mazin, Nature (London) {\bf 464}, 183 (2010).

\bibitem{VASP}
G. Kresse and J. Hafner, Phys. Rev. B {\bf 47}, 558 (1993); G. Kresse
and J. Furthm\"{u}ller, Phys. Rev. B {\bf 54}, 11169 (1996).

\bibitem{SuppMat}
See EPAPS Document No.~1.

\bibitem{MAISE}
A.N. Kolmogorov, http://maise-guide.org.

\bibitem{PAW}
P. E. Bl\"{o}chl, Phys. Rev. B {\bf 50}, 17953 (1994).

\bibitem{MONKHORST_PACK}
J. D. Pack and H. J. Monkhorst, {Phys. Rev. B} {\bf 13}, 5188 (1976);
{\bf 16}, 1748 (1977).

\bibitem{PBE}
J.P. Perdew, K. Burke, and M. Ernzerhof, Phys. Rev. Lett. {\bf 77}, 3865 (1996).

\bibitem{Fe}
L.~Vo\v{c}adlo {\it et al.}, Faraday Discuss. {\bf 106}, 205 (1997).

\bibitem{aB}
M.J.~van~Setten {\it et al.}, J. Am. Chem. Soc. {\bf 129}, 2458 (2007).

\bibitem{PHON}
D. Alf{$\grave{e}$}, {Comp. Phys. Commun.} {\bf 180}, 2622 (2009).

\bibitem{PWscf}
P. Giannozzi {\it et al.}, http://www.quantum-espresso.org.

\bibitem{kq_meshes} 
We employ ultrasoft pseudopotentials [D. Vanderbilt, Phys. Rev. B {\bf
41}, R7892 (1990)] with a cutoff of 43 and 344 Ry for the wave
functions and charge density, respectively. A $9\times 9\times 18$
$k$-point mesh with a Gaussian smearing of 0.02 Ry and a $3\times
3\times 6$ $q$-mesh are used for phonon dispersion calculations; a
$18\times 18\times 36$ $k$-mesh is used to evaluate the $e$-ph
coupling.

\bibitem{TMB}
D. Hohnke and E. Parth\'{e}, Acta Cryst. {\bf 20}, 572 (1966).

\bibitem{bandgap}
G. Onida, L. Reining, and A. Rubio, Rev. Mod. Phys. {\bf 74}, 601 (2002).

\bibitem{FeB2_controversy}
L. Topor and O.J. Kleppa, J. Chem. Thermodynamics {\bf 17}, 1003
(1985); A.F. Guillermet and G. Grimvall, J. Less-Common Met. {\bf 169}, 257 (1991).

\bibitem{LiB2}
A.N. Kolmogorov and S. Curtarolo, Phys. Rev. B {\bf 74}, 224507 (2006) and references therein.

\bibitem{CaC6}
M. Calandra and F. Mauri, Phys. Rev. Lett. {\bf 95}, 237002 (2005).

\bibitem{MgB2_Liu}
A.Y. Liu, I.I. Mazin, and J. Kortus, Phys. Rev. Lett. {\bf 87}, 087005 (2001).

\bibitem{Allen-Dynes}
P.B. Allen and R.C. Dynes, Phys. Rev. B {\bf 12}, 905 (1975).

\bibitem{Pickett}
W.E. Pickett, Physica B {\bf 296}, 112 (2001).

\end{thebibliography}
\end{document}


\title{Supplementary material to: New superconducting and semiconducting Fe-B compounds predicted with an {\it ab initio} evolutionary search}

\author{A.N. Kolmogorov,$^{1}$ S. Shah,$^{1}$ E.R. Margine,$^{1}$ A.F. Bialon,$^{2}$ T. Hammerschmidt,$^{2}$ and R. Drautz$^{2}$}

\affiliation{$^{1}$Department of Materials, University of Oxford, Parks Road, Oxford OX1 3PH, United Kingdom}
\affiliation{$^{2}$Atomistic Modelling and Simulation, ICAMS, Ruhr-Universit\"at Bochum, D-44801 Bochum, Germany}

\date{\today}

\maketitle



{\it Library of structure types} We have considered over 40 commonly
seen $M$-B and $M$-C structure types listed in the ICSD. The list
below is ordered by composition and specifies the Pearson symbol and
the prototype for each entry. 1:12: cF52-UB$_{12}$; 1:7: oI64-MgB$_7$;
3:20: oS46-Na$_3$B$_{20}$; 1:6: cP7-CaB$_6$; 3:14:
tI160-Li$_3$B$_{14}$; 1:4: tP20-UB$_4$, oI10-CrB$_4$, mS10-MnB$_4$;
1:3: tP20-LiB$_3$, hP16-Mo$_{0.8}$B$_3$; 2:5: hR21-Mo$_2$B$_{4.65}$;
1:2: hP3-AlB$_2$, oP6-RuB$_2$, hR18-MoB$_2$, oP12-PbCl$_2$,
hP6-ReB$_2$, hP6-MoS$_2$; 2:3: oS20-V$_2$B$_3$, hP10-Ru$_2$B$_3$; 3:4:
oI14-Ta$_3$B$_4$; 5:6: oS22-V$_5$B$_6$; 1:1: oP8-FeB-b,
oS8-PtB$_{0.67}$, tI16-MoB, oS8-TlI, oP8-TiSi, hP2-WC; 5:4:
hP18-Rh$_5$B$_4$; 11:8: oP38-Ru$_{11}$B$_{8}$; 3:2: tP10-U$_3$Si$_2$,
aP30-Rh$_9$B$_{5.5}$; 5:3: tI32-Cr$_5$B$_3$; 2:1: tI12-CuAl$_2$; 7:3:
hP20-Th$_7$Fe$_3$, oP40-Mn$_7$C$_3$; 5:2: mS28-Mn$_5$C$_2$; 3:1:
oP16-Fe$_3$C, oS16-Re$_3$B, tI32-Ni$_3$P; 23:6: cF116-Cr$_{23}$C$_6$;
4:1: tP10-Be$_4$B. Table \ref{relstruc} summarizes the lattice
parameters and Wyckoff positions of the fully relaxed unit cells of
the relevant ICSD and proposed phases.

{\it Evolutionary algorithm (EA) search} Module for {\it Ab Initio}
Structure Evolution ({\small MAISE}) \cite{MAISE} takes advantage of
the best EA strategies reported in the recent
years \cite{Deaven_Ho,Oganov_GA,USPEX,GA_cuts} as well as our extensive tests
to achieve fastest convergence for metal borides. The module creates
input for {\small VASP} and gets back fully relaxed structures with
the corresponding {\it ab initio} enthalpies. Starting sets of $N=12$
structures are generated randomly. The population is then evolved
through crossover (75\%) and crossover plus mutation (25\%); the
mutation consists of shifts and swaps of the atoms as well as
distortions of the unit cell. The best $N$ members are selected from
the pool of $N$ old and $N$ new members; the probability of survival
is proportional to the fitness depending on the structure's
enthalpy. Special care is taken to detect and eliminate similar
structures in order to avoid stagnation. Fig. 1(a,b) demonstrates that
diversity is maintained throughout the search (grey points) while the
fittest member is always kept in the population (green
points). Fig. 1(c) shows the lowest relative formation enthalpies
obtained in three runs with different random seeds for each
composition. Each run consisted of 50-100 generations, i.e., 600-1200
structural relaxations. Typically, all three runs converged to the
same structure in no more than 60 generations.

{\it Accuracy of the DFT approximations} The 500 eV energy cutoff and
dense $k$-meshes ensure numerical convergence of the formation energy
differences to within a few meV/atom but the systematic errors due to
the approximated form of the xc functionals for the complex Fe-B
system can be expected to be larger. Figure 1 shows dependence of the
relative stabilities for predicted phases on the choice of commonly
used DFT flavours: the Perdew-Zunger parameterization within the local
density approximation (LDA) \cite{LDA} and the Perdew-Burke-Ernzerhof
(PBE) \cite{PBE}, Perdew-Wang (PW91) \cite{PW91}, and
PBEsol \cite{PBEsol} parameterizations within the generalized gradient
approximation (GGA). 

\begin{figure}[t!]
\begin{center}
\includegraphics[width=80mm,angle=0]{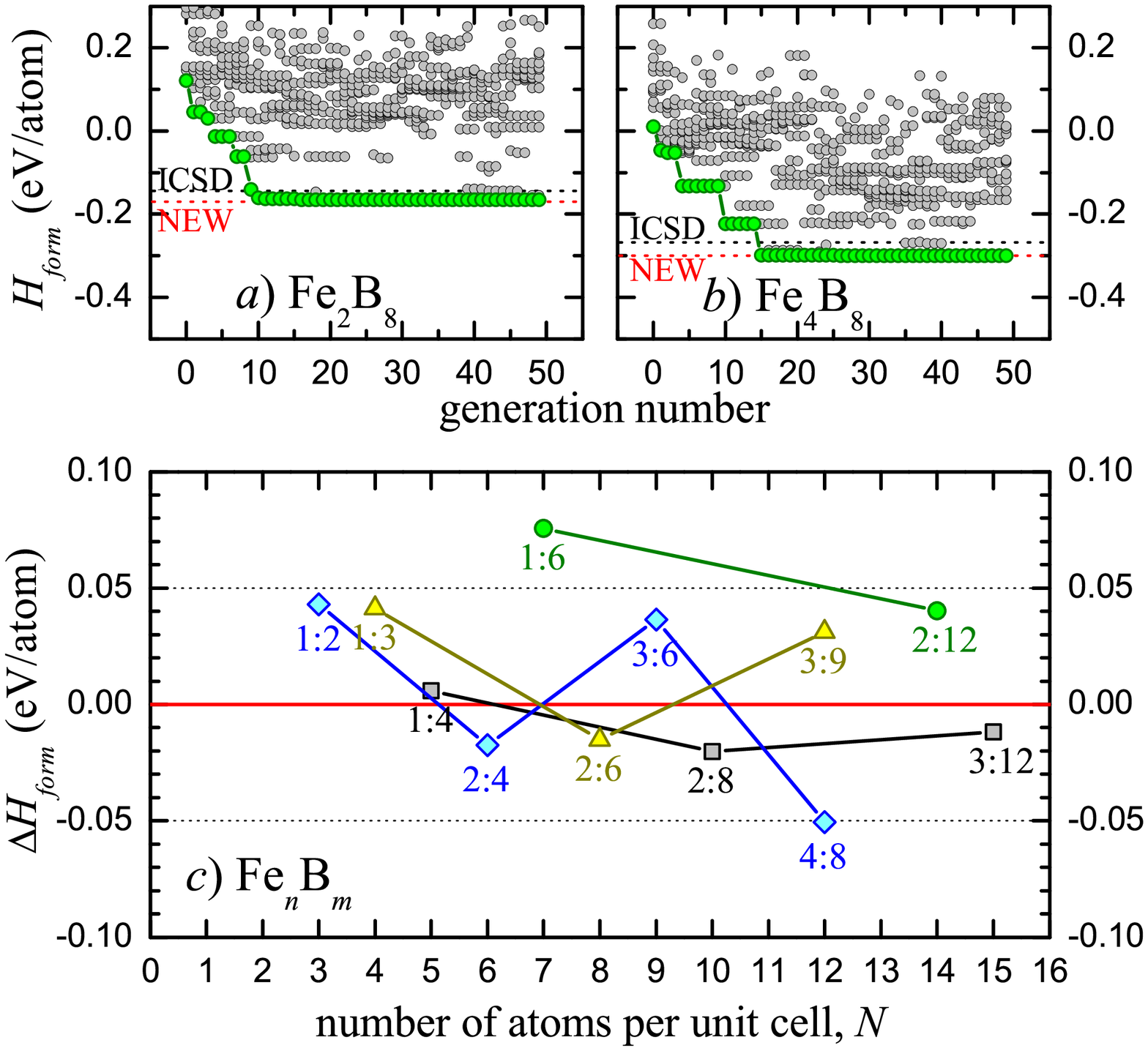}
\caption{ \small {\it a,b}) Typical enthalpy distributions for
competing structures during an EA-driven ground state search. In each
generation 12 new members are fully relaxed with {\small VASP}; the
best member shown in green is kept in the population. For 1:4 and 1:2
Fe-B compositions the evolutionary search finds new structure types,
oP10 and oP12, significantly lower in enthalpy than the known ones
from the ICSD. {\it c}) Relative formation enthalpies (referenced to
the $\alpha$-B$\leftrightarrow$oP8-FeB tie-line) of the most stable
phases for each unit cell size found in the EA search; compounds with
negative $\Delta H_{form}$ are thermodynamically stable w.r.t. the
known $\alpha$-B and oP8-FeB.}
\label{fig5}
\end{center}
\end{figure}

\begin{figure}[t!]
\begin{center}
\includegraphics[width=85mm,angle=0]{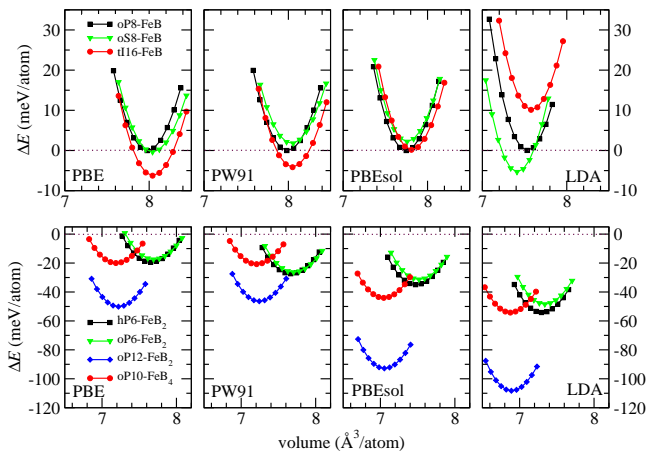}
\caption{ \small E(V) curves for the competing crystal structures
oP8, oS8, and tI16 at 1:1 composition relative to oP8 (upper panel) and
the predicted structures at 1:2 and 1:4 composition (lower panel)
relative to the $\alpha$-B$\leftrightarrow$oP8-FeB tie-line.}
\label{fig6}
\end{center}
\end{figure}
We observe that the relative stability of the boron-rich phases varies
by over 30 meV/atom when simulated in GGA-PBE and LDA but in all the
treatments the predicted FeB$_2$ and FeB$_4$ compounds are stable
w.r.t. the $\alpha$-B$\leftrightarrow$oP8-FeB tie-line. The systematic
errors do not allow us to conclusively resolve the close in energy
oP6-FeB$_2$ and hP6-FeB$_2$ or tI16-FeB, oP8-FeB and oS8-FeB. However,
oP10-FeB$_4$ is consistently below oI10-FeB$_4$ by $\sim25$ meV/atom
and hP3-FeB$_2$ is consistently above the
$\alpha$-B$\leftrightarrow$oP8-FeB tie-line by $\sim200$~meV/atom. The
thermodynamic corrections are found to be not sensitive to the choice
of the DFT approximation and show similar $T$-dependences in the LDA
and GGA-PBE.

{\it Convergence of the $e$-ph calculations} In order to check the
convergence of $\lambda$ we perform the calculations for two sets of
grids: (i) the $2\times2\times4$ $q$- and $6\times6\times12$
$k$-meshes for the dynamical matrix and the $12\times12\times24$
$k-$mesh for the $e$-ph coupling (ii) the $3\times3\times6$ $q$- and
$9\times9\times18$ $k$-meshes for the dynamical matrix and the
$18\times18\times36$ $k$-mesh for the $e$-ph coupling. For the 0.02 Ry
smearing in the $k$-point integration we find nearly identical values
of $\lambda=0.80$ (note that in these calculations the smearing leads
to a lower value of $n(E_F)\approx0.8$ states/(eV spin f.u.) due to a
large gradient of the DOS near the Fermi level). We also find
negligible changes (below 0.4 cm$^{-1}$) in $\Gamma$-point frequencies
when the $k$-mesh is increased from $9\times9\times18$ to
$12\times12\times24$ and $15\times15\times30$.
\begin{widetext}
\begin{center}
 \begin{table*}[h!]
 \hfill{}
    \begin{tabular}[t]{ccccccccccccccc}\hline\hline
                     &  & Space   & &   $a$  &  $b$  &   $c$  &$\beta$ & &\multicolumn{5}{c}{Wyckoff positions}                               \\
                     &  & group   & & (\AA)  & (\AA) &  (\AA) &        & &\multicolumn{2}{c}{B }       & & \multicolumn{2}{c}{Fe}             \\\hline
       oP8-FeB       &  &  62     & &  5.409 & 2.948 &  3.997 &        & & 4$c$ & (0.0344,  1/4  , 0.6186) & & 4$c$ &(0.1779,  1/4  , 0.1203) \\
       oS8-FeB       &  &  63     & &  2.896 & 7.521 &  2.949 &        & & 4$c$ & (  0   , 0.4306,  1/4  ) & & 4$c$ &(  0   , 0.1429,  1/4  ) \\
       tI16-FeB      &  & 141     & &  2.931 & 2.931 & 14.971 &        & & 8$e$ & (  0   ,  1/4  , 0.7151) & & 8$e$ &(  0   ,  1/4  , 0.1791) \\
       hP3-FeB$_2$   &  & 191     & &  3.030 &       &  2.749 &        & & 2$c$ & ( 1/3  ,  2/3  ,   0   ) & & 1$b$ &(  0   ,   0   ,  1/2  ) \\
       oP6-FeB$_2$   &  &  59     & &  4.411 & 2.778 &  3.764 &        & & 4$f$ & (0.0467,  1/4  , 0.6428) & & 2$a$ &( 1/4  ,  1/4  , 0.1504) \\
       hP6-FeB$_2$   &  & 194     & &  2.805 &       &  6.737 &        & & 4$f$ & ( 1/3  ,  2/3  , 0.5567) & & 2$c$ &( 1/3  ,  2/3  ,  1/4  ) \\
       oP12-FeB$_2$  &  &  62     & &  4.816 & 4.807 &  3.740 &        & & 8$d$ & (0.1611, 0.0693, 0.3930) & & 4$a$ &(0.0198,  1/4  , 0.8761) \\
       mP8-FeB$_3$   &  &  11     & &  3.813 & 2.819 &  5.527 &100.974 & & 2$e$ & (0.1766,  1/4  , 0.0596) & & 2$e$ &(0.9137,  1/4  , 0.6891) \\
                     &  &         & &        &       &        &        & & 2$e$ & (0.3759,  1/4  , 0.5476) & &      &                         \\
                     &  &         & &        &       &        &        & & 2$e$ & (0.4872,  1/4  , 0.8831) & &      &                         \\
       oI10-FeB$_4$  &  &  71     & &  4.518 & 5.287 &  3.052 &        & & 8$n$ & (0.2121, 0.3407,   0   ) & & 2$a$ &(  0   ,   0   ,   0   ) \\
       oP10-FeB$_4$  &  &  58     & &  4.521 & 5.284 &  3.006 &        & & 4$g$ & (0.1606, 0.1267,   0   ) & & 2$c$ &(  0   ,  1/2  ,   0   ) \\
                     &  &         & &        &       &        &        & & 4$g$ & (0.7508, 0.1871,   0   ) & &      &                         \\
       mS30-FeB$_4$  &  &  12     & & 13.961 & 2.936 &  5.316 & 96.256 & & 4$i$ & (0.7515,   0   , 0.7495) & & 2$b$ &(  0   ,  1/2  ,   0   ) \\
                     &  &         & &        &       &        &        & & 4$i$ & (0.1124,   0   , 0.1584) & & 4$i$ &(0.1610,   0   , 0.5467) \\
                     &  &         & &        &       &        &        & & 4$i$ & (0.9164,   0   , 0.1609) & &      &                         \\
                     &  &         & &        &       &        &        & & 4$i$ & (0.7864,   0   , 0.0661) & &      &                         \\
                     &  &         & &        &       &        &        & & 4$i$ & (0.4481,   0   , 0.3587) & &      &                         \\
                     &  &         & &        &       &        &        & & 4$i$ & (0.5817,   0   , 0.3346) & &      &                         \\\hline\hline
    \end{tabular}
\hfill{}
    \caption{Space groups, lattice parameters, and Wyckoff-positions
      of fully relaxed Fe-B structures.}
    \label{relstruc}
\end{table*}
\end{center}
\end{widetext}
%